**Original Research**

# Effects of Gamma Ray Bursts in Earth's Biosphere

Osmel Martin[1], Rolando Cardenas[2], Mayrene Guimaraes[3], Liuba Peñate[4], Jorge Horvath[5], Douglas Galante[6]

[1,2] *Department of Physics, Universidad Central de Las Villas, Santa Clara, Cuba.*
*Phone 53 42 281109 Fax 53 42 281109. e-mail:* [1] osmel@uclv.edu.cu, [2] rcardenas@uclv.edu.cu

[3] *Marine Ecology Group, Center for Research of Coastal Ecosystems, Cayo Coco, Ciego de Avila, Cuba*
*Phone 53 33 301104 ext 117 e-mail:* mayrene@ciec.fica.inf.cu

[4] *Department of Biology, Universidad Central de Las Villas, Santa Clara, Cuba.*
*Phone 53 42 281109 Fax 53 42 281109 e-mail:* liubapa@uclv.edu.cu

[5, 6] *Department of Astronomy, Instituto de Astronomia, Geofísica e Ciências Atmosféricas, Universidade de São Paulo, São Paulo, Brazil. e-mail:* [4] foton@astro.iag.usp.br; [5] douglas@astro.iag.usp.br

**Abstract:** We continue former work on the modeling of potential effects of Gamma Ray Bursts on Phanerozoic Earth. We focus on global biospheric effects of ozone depletion and show a first modeling of the spectral reduction of light by $NO_2$ formed in the stratosphere. We also illustrate the current complexities involved in the prediction of how terrestrial ecosystems would respond to this kind of burst. We conclude that more biological field and laboratory data are needed to reach even moderate accuracy in this modeling.

## 1 Introduction

The idea of strong astrophysical influence in the course of the Earth's biological evolution has been discussed by several authors. According with that, large asteroid impacts, giant solar flares, supernovae explosions (SNe) or Gamma Ray Bursts (GRB's), could have acted as triggers of extinctions in the Earth's geological past. For instance, recently it has been suggested a connection between supernovae and the extinction of tropical American mollusks that took place around the Pliocene - Pleistocene boundary (Benitez, Maiz-Apellaniz & Canelles, 2002**).** Additionally, the potential influence of SNe is used to define concepts such as Galactic Habitable Zone, in a more astrobiological context (Gonzalez, Brownlee & Ward, 2001).

For the case of transient radiation events like SNe or GRB's, beyond differences between them, their main influences on the Earth's biota could be similar. Both events are strong sources of highly energetic gamma radiation, capable of inducing severe perturbations on the chemistry of planetary atmospheres. Their main effects on the Earth's biota are strongly dependent on the atmospheric composition, the presence or not of an active $O_2/O_3$ ultraviolet radiation (UVR) blocking system, and ecosystem specificities. They are associated to an increase of typical UV levels reaching the ground at least in two forms: the so called reemitted UV flash and the increase of solar UV by depletion of the ozone layer (Galante & Horvath, 2007). The relative importance of these effects appears to be a strong function of the free oxygen content in the atmosphere.

For contemporary Earth-like atmospheres (rich in $O_2$), the main influence is the depletion of the ozone layer through the catalytic effect of $NO_x$ species formed during the burst. The total recovery is determined mainly by the atmospheric chemistry and the transport processes. The time for the recovery of the ozone is around a decade (Thomas et al, 2005). Other potential influences on the climate and biosphere may be induced by abnormal nitrate deposition due to rainout of $NO_x$ in the form of nitric acid rain or cooling effects and reduction of sunlight in the visible range due to high $NO_2$ levels.

Modeling the action of an UVR excess on Earth's biosphere is a highly complicated task, given the variability in species sensitivity, possible thresholds, and non-linearities. We can restrict our consideration to the so called primary producers of the biosphere (phytoplankton, algae, higher plants), as they form the basis of the food web, so any perturbation on them should be reflected in higher levels of the trophic assemblage (herbivores, carnivores, omnivores). As the biosphere contributes to the $CO_2$ fixation and $O_2$ evolution, important perturbations on it by an UVR excess, coming from any source (solar or extrasolar), have the potential for global climate changes. However, the biosphere is formed by many interacting ecosystems, whose respective responses to UVR excess is even today an open question (Thomas, 2008; Hader et al, 2007).

In this work, we continue exploring global and regional effects that a GRB could cause on Earth's biosphere, due to the aforementioned ozone depletion and enhanced atmospheric opacity due to the formation of $NO_2$ as consequences of the burst impact.

## 2 Basic assumptions

### 2.1 <u>The effects on the atmosphere</u>

Ionizing radiation dissociates $N_2$ and $O_2$ in the atmosphere, releasing important quantities of atomic nitrogen and oxygen. These very reactive chemical species then form considerable quantities of nitrogen oxides, catalizers of the ozone dissociation. As stated in (Thomas et al. 2005), the ''typical'' nearest burst in the last billion years would cause a globally averaged ozone depletion of up to 38% and significant global depletion (at least 10%) would persist up to seven years. This would imply:

- an enhanced irradiation of the planet's surface with the solar ultraviolet radiation (UVR),
- atmospheric opacity reducing visible sunlight in a few percent because of the formation of $NO_2$, with potential global cooling, and,
- deposition of nitrate through rainout of nitric acid, slightly greater than that currently caused by lightning, lasting several years.

In order to account for the spectral reduction of irradiance at planet surface due to the formation of $NO_2$ we used the solar spectrum $I_0(\lambda)$ at surface as given in (ASTM G173 - 03e1). Then, considering that in (Thomas et al. 2005) total irradiance reduction in the range (0-10) % due to the formation of $NO_2$ is reported, we calculated which columns of $NO_2$ would make reductions of total irradiance $I$ given by several values of the fraction number $f$:

$$\frac{I_{after}}{I_{before}} = f \qquad (1)$$

where from now on the subscripts *after* and *before* mean after and before the impact of the GRB. We used the values for $f$ of 0.98, 0.96, 0.94 and 0.92, representing irradiance reductions of 2, 4, 6 and 8 % respectively.

The values of total irradiances after and before the burst are given by:

$$I_{before} = \int_{280\ nm}^{700\ nm} I_0(\lambda) d\lambda \qquad (2)$$

$$I_{after} = \int_{280\ nm}^{700\ nm} I_0(\lambda) e^{-\tau} d\lambda \qquad (3)$$

where $\tau$ is the optical path of photons in the $NO_2$ column. This magnitude gives the clue to estimate the quantity of $NO_2$ needed to reduce the total irradiance in a given *f*. The above procedure neglects the increase of irradiance due to ozone depletion, but as the Sun peaks in the visible part of the spectrum, that contribution to the total irradiance $I_{after}$ is very small, something that we checked using the radiative transport code NCAR/ACD TUV: Tropospheric Ultraviolet & Visible Radiation Model (NCAR/ACD).

2.2 Estimation of wide-scale damage on the biosphere

It is clear that the first and second atmospheric effects mentioned above could affect many photosynthetic species: more solar UV can damage DNA and inhibit photosynthesis to some extent, while less visible sunlight (i. e., photosynthetic active radiation, PAR) would reduce the energy available for photosynthesis and therefore for primary production.

However, the third effect can offset, at least partially, the above mentioned inhibition of photosynthesis, and could even cause eutrofication (over-enrichment of nutrients) in some freshwater and coastal ecosystems. It is true that the nitric acid rain could stress portions of the biosphere, but, after titration, the increased nitrate deposition could be helpful to photosynthetic organisms, especially to land plants. This effect requires further attention and is not a focus in this paper.

Due to the considerable variability in species sensitivity to radiations and to non linearities, the accurate modeling of how the biosphere would behave in excess of UVR is very complicated. However, a rough idea of the biological effects of ozone depletion is the radiation amplification factor (RAF), relating the biological effective irradiances $E*$ with the ozone columns *N*, after and before the ionizing event:

$$\frac{E^*_{after}}{E^*_{before}} = \left(\frac{N_{before}}{N_{after}}\right)^{RAF} \qquad (4)$$

The RAF's are dependent both upon the group of species and upon the organismal process to be considered (represented by a biological weighting function BWF).

BWF's are typically measured in controlled laboratory conditions; so they are of limited value to estimate the actual response of living beings to UVR. Under the action of UVR, organisms can enzymatically reverse the photochemical reaction or re-synthesize the affected molecules. These processes, generically known as repair, depend not only on the species, but also on environmental variables. For instance, it is well known the interaction repair – temperature for several species of phytoplankton: at very low temperatures repair is very slow, while at intermediates temperature repair is good. In general, repair is not properly taken into account when BWF's are measured; therefore the biological amplification factor (BAF) is the quantity which would give us more accurate information on the biological effects of UVR:

$$\frac{P_{after}}{P_{before}} = BAF \times \frac{E^*_{before}}{E^*_{after}} \qquad (5)$$

where $P$ is the rate of an organismal process (for example, photosynthesis).

Unfortunately, very few BAF's have been measured, though the alternative exposure - response curves (ERC's) for several species have been reported. Anyway, RAF's and BAF's could be useful for a first rough approach to estimate global damage on the biosphere of a Gamma Ray Burst, but a more detailed modeling implies to look at specific ecosystems, the building blocks of the biosphere.

### 2.3 Gamma Ray Bursts at ecosystem level

From the broadest biophysiological point of view, the biosphere is the global ecological system integrating all living beings and their relationships, including their interaction with the elements of the lithosphere, hydrosphere, and atmosphere. The biosphere can also be considered as the sum of all ecosystems (aquatic, terrestrial and hybrid). Studies of ecosystems usually focus on the movement of energy and matter through the system, but these processes will depend on the kind of ecosystem. However, some generic characteristics can be stated:

- On energy: Almost all ecosystems run on energy captured from the Sun by primary producers (phytoplankton, algae, higher plants) via photosynthesis, this energy then flows through the food chains to primary consumers (herbivores, who eat and digest the plants), and on to secondary and tertiary consumers (either carnivores or omnivores).
- On matter: It is incorporated into living organisms by the primary producers. Photosynthetic plants fix carbon from carbon dioxide and nitrogen from atmospheric nitrogen or nitrates present in the soil to produce amino acids. Much of the carbon and nitrogen contained in ecosystems is created by such plants, and is then consumed by secondary and tertiary consumers and incorporated into themselves. Nutrients are usually returned to the ecosystem via decomposition. The entire movement of chemicals in an ecosystem is termed a biogeochemical cycle, and includes the carbon and nitrogen cycle.

To study the effects of GRB's at regional or local level implies modeling the action of UVR excess on several different ecosystems. In this work we have chosen lakes, one of the reasons being that the selected model of lake successfully describes the process of eutrofication (over enrichment by nutrients, primarily nitrogen and phosphorus) and it has been predicted that one of the atmospheric effects of a GRB would be an increased rainout of nitrogen compounds, thus contributing to the eutrofication of terrestrial ecosystems (Thomas et al 2005). We expect then that our results might be a rough proxy of what could happen in a considerable proportion of inland waters and coastal ecosystems after the incidence of the UVR perturbation, because many of these systems often show some degree of eutrofication due to the influence of land masses. We admit that a more accurate modeling of the action of an UVR excess at ecosystem level would require specific models for other specific ecosystems, both aquatic and terrestrial, something which we leave for future work.

#### 2.3.1 The Comprehensive Aquatic Simulation Model

The Comprehensive Aquatic Simulation Model (CASM) has successfully described the key features of the eutrophication process in real lakes (Amemiya et al, 2007). This process is associated to the over enrichment by nutrients, primarily phosphorus and nitrogen, with a consequent increase of phytoplankton levels, while other species such as fish and zooplankton become rather scarce. As we said in the above subsection, eutrofication by nitrate deposition is one of the potential consequences of a GRB striking our atmosphere (Thomas et al, 2005), making this model attractive for our purposes. In this model there is an external input $I_N$ of the limiting nutrient $N$ to the ecosystem, which in our case would include the atmospheric deposition of nitrates after the GRB by rainout. Equation (6) below represents the dynamics of nutrients in the ecosystem, where $r_N$ is the loss rate of nitrogen by diverse causes (for instance, sedimentation, flow out, etc.), while the third term of right hand side (rhs) models the consumption of nutrients by the primary consumers (phytoplankton $X$). The form of this term is inspired in the Michaelis- Menten kinetics, firstly applied to simple processes in which enzymes participate. In our case, $\gamma$ is the ratio of nutrient mass (nitrate mass) to biomass, $r_1$ is the maximum growth rate of phytoplankton and $k_1$ is a half saturation constant (when $N = k_1$, the whole term will be divided by two after cancelling $N$, hence the denomination *half saturation*). Finally, the fourth term of rhs of eq. (6) represents the input of nutrient $N$, via decomposition of detritus matter $D$, considering that $d_4$ is the decomposition rate of $D$.

$$\frac{dN}{dt} = I_N - r_N - \frac{\gamma r_1 NX}{k_1 + N} + \gamma d_4 D \qquad (6)$$

The primary production of the ecosystem is represented by equation (7) below, where phytoplankton $X$ consumes nutrients via the first term of rhs (compare it with the third term of rhs on the equation (6)), and the second term shows how zooplankton $Y$ predates on phytoplankton. In this term, $f_1$ is the feeding rate of zooplankton and $k_2$ is the half saturation constant for this term (because when $X^2 = k_2$, the cancellation of $X^2$ ensures that the whole term is divided by two). The last term of rhs of the equation contains the mortality $d_1$ of phytoplankton and its removal rate from the ecosystem $e_1$.

$$\frac{dX}{dt} = \frac{r_1 NX}{k_1 + N} - \frac{f_1 X^2 Y}{k_2 + X^2} - (d_1 + e_1)X \qquad (7)$$

Equation (8) below represents the dynamics of the primary consumer, zooplankton $Y$. The first term of rhs shows how it predates on phytoplankton (compare it with the second term of rhs of above equation), while the third term says how zooplankton is eaten by the secondary consumer, the zooplanktivorous fish $Z$. The parameter $\eta$ represents the assimilation efficiency of zooplankton, the meanings of the other parameters can readily be deduced from the explanations given of the first two equations.

$$\frac{dY}{dt} = \frac{\eta f_1 X^2 Y}{k_2 + X^2} - \frac{f_2 Y^2 Z}{k_3 + Y^2} - (d_2 + e_2)Y \qquad (8)$$

The dynamics of the secondary consumer, the zooplanktivorous fish, is given by equation below. Here the new parameter $Z^*$, the low equilibrium biomass of zooplanktivorous fish, avoids the unrealistic situation of former versions of CASM, in which fish could appear from states in which it was already extinct.

$$\frac{dZ}{dt} = \frac{\eta f_2 Y^2 Z}{k_1 + Y^2} - (d_3 + e_3)(Z - Z^*) \qquad (9)$$

Finally, we should consider that there are sources of detritus matter $D$ in the ecosystem (fecal material and dead $X$, $Y$ and $Z$), whose decomposition returns nutrients to the ecosystem. This is very important in all ecosystems: an important fraction of nutrients is returned to the ecosystem via decomposition of feces and dead beings, as stated in equation below:

$$\frac{dD}{dt} = \frac{(1-\eta) f_1 X^2 Y}{k_2 + X^2} + \frac{(1-\eta) f_2 Y^2 Z}{k_{3_1} + Y^2} + d_1 X + d_2 Y + d_3 Z - (d_4 + e_4) D, \qquad (10)$$

We remind that $d_i$ are death or decomposition rates and $e_i$ are removal rates from the system. As can be seen from equations (6)-(10), CASM has five dynamical variables and 19 parameters. In general, we refer the interested reader to (Amemiya et al, 2007) for more details.

2.3.2 The inclusion of radiative transport in the Comprehensive Aquatic Simulation Model

The formulation of CASM model above does not take into consideration the vertical distribution of the living species in the water column. This is an important omission when considering any situation of UVR stress, given the attenuation of radiation due to the phenomena of absorption and dispersion in the water column. To account for this we considered phytoplankton to be the only trophic level stressed by UVR, as they are obligated to have an adequate solar exposure in order to perform photosynthesis. We can then imagine all phytoplankton living in an effective depth and receiving increased UVR levels after a GRB. Thus, to include the role of some components of the ecosystem as UV screeners in the water column (detritus and phytoplankton themselves), we modified the CASM model considering the mortality rate coefficient of phytoplankton ($d_1$) no longer a constant, but an explicit function of such components of the form

$$d_1 = e^{-h_X X - h_D D} d \qquad (11)$$

The above exponential dependence is motivated by the well known Beer's law for the absorption of light by any liquid solution, $h_X$ and $h_D$ are coefficients for UVR attenuation by phytoplankton and detritus matter, while $d$ is the lethality rate coefficient of the phytoplankton when no UV blocking effect is considered.

## 3 Results and General Discussion

3.1 Global damage: the biosphere level

3.1.1 The effects of ozone depletion

As mentioned above, in (Thomas et al. 2005) it is shown that the typical nearest burst in the last billion years would cause an averaged global ozone depletion of up to 38%, which would persist several years. For instance, seven years after the burst, 10% ozone depletion would be expected. Considering this, in Table 1 below we show the fractional increase of the effective biological irradiances for several values of ozone depletion and several biological weighting functions.

| Biological Weighting Function | RAF | $\frac{E^*_{after}}{E^*_{before}}$ for several values of ozone depletion (%) | | | |
|---|---|---|---|---|---|
| | | 38 | 30 | 20 | 10 |
| Photoinhibition of a marine phytoplankton | 0.31 | 1.16 | 1.12 | 1.07 | 1.03 |
| Photoinhibition of land plants | 0.51 | 1.27 | 1.20 | 1.12 | 1.05 |
| DNA damage | 1.67 - 2.2 | 2.22 - 2.85 | 1.82 – 2.20 | 1.45 – 1.63 | 1.19 – 1.26 |

**Table 1** Radiation Amplification Factors and fractional increase of effective biological irradiances for some biological weighting functions and for ozone depletions of 38, 30, 20 and 10 %

The table above suggests that DNA damage is in general the main influence of a GRB over the biosphere and that land plants might suffer more than phytoplankton. However, it should be noticed that RAF's are typically measured in controlled conditions very different from the natural conditions in which organisms live. Therefore, the use of biological amplification factors (BAF's) or exposure response curves (ERC's) should give us a much better picture of the response of the biosphere to UVR perturbations. Unfortunately, very few BAF's or ERC's have been measured for the most common primary producers in the biosphere, such as the main species of marine phytoplankton. Therefore, we are lacking biological field data to make more accurate accounts of the potential global effects of a GRB on the biosphere. The good news is that several studies are now underway which will supply useful biological data, therefore the next future looks promissory.

3.1.2 The effects of irradiance reduction due to $NO_2$ formation

We followed the methodology explained in subsection 2.1 to calculate the spectral reduction of light as a consequence of the enhanced formation of $NO_2$. The Table 2 below shows a slightly selective absorption in the visible band (PAR), while a more pronounced absorption in the UV-A band and in the photorepair band (350-450nm) appears. The photorepair light is needed to execute the most efficient repair pathway of DNA damages caused by UV-B.

| $f$ | $f_{UV\text{-}A}$ | $f_{PAR}$ | $f_{350\text{-}450\ nm}$ |
|---|---|---|---|
| 0.98 | 0.92 | 0.98 | 0.92 |
| 0.96 | 0.85 | 0.95 | 0.84 |
| 0.94 | 0.78 | 0.93 | 0.77 |
| 0.92 | 0.71 | 0.90 | 0.70 |

**Table 2** Ratio of irradiances after and before the burst (eq. 1), both for global irradiance and for some bands

We additionally checked that 30% depletion of the standard ozone column of 340 Dobson units implied a 22% increase of UV-B, but only a 0.37 % increase of UV-A, therefore the ozone depletion contribution to the increase of UV-A is much smaller than the decrease of this band due to $NO_2$ formation, which in this case depletes light in around 10 %.

Thus, the net global biological effect of a GRB suggests a combination of more damages due to more UV-B reaching the ground (because of ozone depletion) and a less efficient repair of DNA damages because less light in the photorepair band (350nm–450 nm) reaches the ground. Also, less light (PAR) would be available for photosynthesis. Additionally, the total reduction of sunlight in the percents stated in this work has the potential of global cooling, something which per se deserves considerable future investigation.

3.2 Regional damage: the ecosystem level

As stated in subsection 2.3, to take into account the combined effects of the depletion of the ozone layer and the (spectral) reduction of sunlight, our modification of the CASM model for lakes was explored with increments of the mortality rate coefficient of phytoplankton ($d_1$).

In Figure 1 it is shown how the qualitative behavior of the model changes as a function of the parameter ($d_1$).

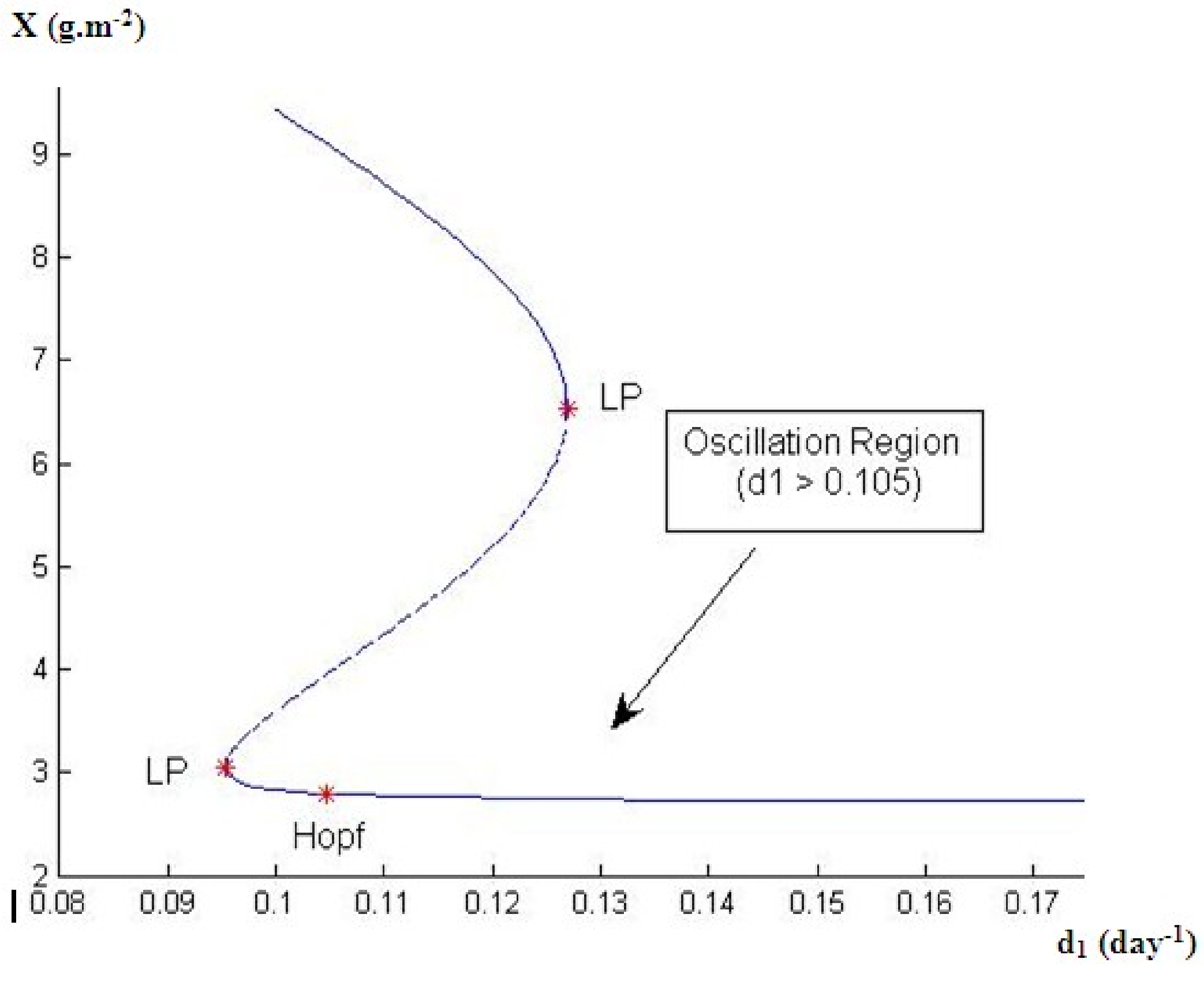

**Fig. 1:** Bi-stability appears for mortality rates of phytoplankton ($d_1$) in the approximate range (0.095 – 0.125) $day^{-1}$ (solid lines represent stable or oscillatory states, dashed

lines transient ones). The Hopf bifurcation at $d_1$= 0.105 $day^{-1}$ marks the transition from a stable state to an oscillatory one.

When the parameter $d_1$ increases to $d_1$=0.105, only a 5 percent above the referenced value $d_1$= 0.1 in (Amemiya et al, 2007), the steady clear state emerges as an oscillating state. At higher values (around $d_1$ = 0.125), the bi-stability of the system is broken and the oscillating state emerges as a unique possibility. Such alternative states are exhibited by CASM for other parameter regions (Amemiya et al, 2007).

Radiative transport analysis in oscillating regimes appears interesting because the optical properties of the water column are continually varying in the time. Some components as detritus matter (D) and phytoplankton (X) play additional UV protection to the main underwater species. Taking into account our modified expression for the mortality rate coefficient (eq. 11) and equal contributions to the attenuation of UV photons by phytoplankton and by detritus ($h = h_X = h_D$), we found the behavior shown in Figure 2.

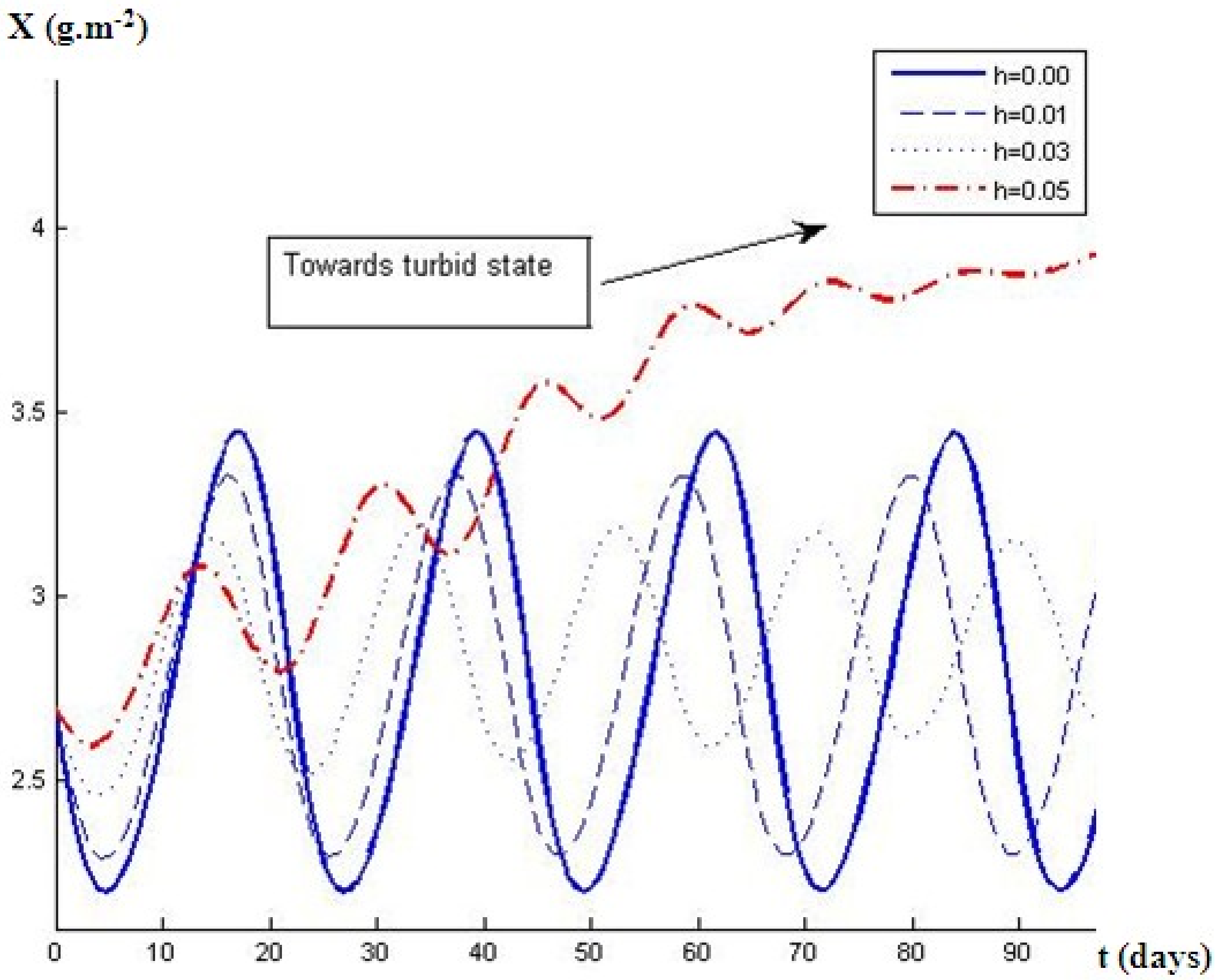

**Fig. 2** Self protection from UV photons of detritus and phytoplankton could cause regime shift or oscillations around the clear state

Now, according with the information in Figure 2, if self-protection is not too high, the oscillating regime around the clear state persists, with minor corrections in the amplitude and oscillation period. If the self-protection reaches some threshold value, phytoplankton population recoveries progressively in time and the ecosystem comes back to the original turbid state.

The above results illustrate the complexities involved in predicting how terrestrial ecosystems would recover if stressed by a GRB. The status towards which a given lake will evolve might depend on several variables and parameters, but phytoplankton, being the primary producer, would play the determinant role. However, a more accurate modeling of the recovery of aquatic ecosystems after a GRB needs a closer look at the behavior of the more common species of phytoplankton under UVR stress, and also other environmental variables are probably to be taken into consideration.

## 4 Conclusions

Given non-linearity and variability in the response of biological systems to radiations, it is difficult to predict the damage and recovery of the biosphere under the impact of the ''typical'' nearest GRB in the last billion years. In this work we have estimated some global effects on the biosphere, but the lack of data on biological amplification factors for the more abundant species of primary producers actually limit the predictive power of present studies. However, largely motivated by today's ozone depletion, some researchers are currently making studies on the response of the most abundant primary producers to UVR; therefore soon we should be able to make more detailed modeling on the potential global biological effects of a GRB.

On the ecosystem (regional) scale a similar situation holds, but again we are optimist concerning the next arrival of new field data. This could serve as a discriminating tool to reveal towards which state several terrestrial ecosystems would shift their equilibrium after the action of a nearby GRB.

## Acknowledgments

The authors thank to the Brazilian federal organization CAPES for financial support of the project "Influence of cosmic radiations on Earth's environment and biosphere" and to the Cuban Ministry for Science, Technology and Environment for funding our research project ''Mathematical Modeling of Ecosystems''.